# Shear-induced reaction-limited aggregation kinetics of Brownian particles at arbitrary concentrations

Alessio Zaccone, Daniele Gentili, Hua Wu (吴华), and Massimo Morbidelli

Department of Chemistry and Applied Biosciences

ETH Zurich, 8093 Zurich, Switzerland

**ABSTRACT**

The aggregation of interacting Brownian particles in sheared concentrated suspensions is an important issue in colloid and soft matter science *per se*. Also, it serves as a model to understand biochemical reactions occurring in vivo where both crowding and shear play an important role. We present an effective medium approach within the Smoluchowski equation with shear which allows one to calculate the encounter kinetics through a potential barrier under shear at arbitrary colloid concentrations. Experiments on a model colloidal system in simple shear flow support the validity of the model in the range considered. By generalizing Kramers' rate theory to the presence of collective hydrodynamics, our model explains the significant increase in the shear-induced reaction-limited aggregation kinetics upon increasing the colloid concentration.

## I. INTRODUCTION

The kinetics of coagulation (i.e. irreversible aggregation) and cluster-growth in dilute, stagnant colloidal systems is fairly well understood, such that theoretical models can accurately predict experimental data under most conditions of practical interest.[1] This is not true in the case of colloidal systems under shear, especially under *nondilute* conditions. A theoretical understanding of colloidal stability and aggregation kinetics under shear is very much in demand in view of the crucial role that these phenomena play in the dynamics of biofluids, which are constantly exposed to shear flow (relevant examples are the shear-induced aggregation of platelets in hemostatic processes,[2] of the proteins constituting the synovial fluid,[3] as well as of proteins involved in fibrillogenesis[4], and in pharmaceutical processes[5]). In particular, charge-stabilized suspensions under shear are commonly observed to exhibit quite bizarre colloidal stability: they can remain perfectly stable over extended periods of time and then suddenly jam into solid-like pastes. This behaviour may cause very significant losses in the industrial handling of disperse materials (e.g. in the polymer industry) and is responsible for the deterioration of pigments.[6] On the other hand, there are also situations where shear-induced aggregation and jamming are an essential step in the formation of interesting materials: this is the case of spider-silk (a material with exceptional mechanical properties) formed as a result of shear-induced coagulation of protein aggregates in the spider spinneret.[7] In all these cases, the shear-induced aggregation kinetics is initially very slow until it "explodes" becoming extremely fast.[4-8] The physics underlying this phenomenon has been explained recently in terms of a barrier-hopping process (where the interaction barrier is due to charge-stabilization) driven by the shear.[9] The two-body theory is able to account for the observed induction delay and exponential dependence of the characteristic aggregation time on the shear rate.[9] However, two-body theory applies rigorously only in the limit of infinite dilution ($\phi \to 0$), and neglects important density-dependent effects. This is a strong limitation to the applicability of the theory to e.g. biochemical and biological reactions in vivo which usually take place in crowded environments.[10]

In this work we propose a theoretical model for the characteristic time of shear-induced reaction-limited aggregation at arbitrary colloid volume fraction (relevant for both biological and industrial applications) by generalizing the Smoluchowski problem with shear through an effective medium approach which allows us to account for the effects of colloid concentration such as collective hydrodynamics.

## II. DERIVATION

Let us recall that the governing equation for the steady-state in the pair-correlation function $c(\mathbf{r}) = c_0 g(\mathbf{r})$ (where $c_0$ is the bulk concentration) is the following two-body Smoluchowski equation with convection[9,11]

$$\nabla \cdot \{\beta D[-\nabla U(r) + b\mathbf{v}(\mathbf{r})] - D\nabla\} c(\mathbf{r}) = 0 \qquad (1)$$

where $D = 2D_0 \mathcal{G}(r)$ ($D_0$ being the self-diffusion coefficient and $\mathcal{G}(r)$ the hydrodynamic correction for viscous retardation), $\mathbf{v}(\mathbf{r})$ is the imposed fluid velocity field, $b = 3\pi\mu a$ is the friction coefficient on a tagged particle (with $\mu$ the solvent viscosity and $a$ the colloid radius), and $U(r)$ is the isotropic pair-interaction potential. For concentrated systems, Eq. (1) should be rewritten by taking into account that the friction that an individual particle experiences under *nondilute* conditions has an additional contribution from the hydrodynamic interactions transmitted by the other Brownian particles. To this aim, we adopt here an *effective medium* approach where the solvent is replaced with an *effective* fluid having the macroscopic properties of the suspension.[11] We start by introducing an effective friction coefficient $b_{eff}$ (see Dhont,[11] pp. 356-357) defined via the following Einstein relation

$$D_{eff} = (\beta b_{eff})^{-1} \mathcal{G}(r) = D_{eff}^{\infty} \mathcal{G}(r) \tag{2}$$

where $D_{eff}$ is the relative (long-time) diffusion coefficient for two particles embedded in the effective fluid. Within this approach, the effective friction coefficient is equal to[11]

$$b_{eff} = 3\pi\eta(\phi)a \tag{3}$$

where $\eta$ is the concentration-dependent *effective* (macroscopic) viscosity of the suspension. We can thus rewrite the Smoluchowski equation with shear for two interacting Brownian particles moving in an effective medium representing the suspension as

$$\nabla \cdot \left\{ \beta D_{eff}(\phi)[-\nabla U(r) + b_{eff}(\phi)\mathbf{v}(\mathbf{r})] - D_{eff}(\phi)\nabla \right\} c(\mathbf{r}) = 0 \tag{4}$$

where $D_{eff}$ is given by Eqs. (2)-(3) and is now also a function of the colloid volume fraction $\phi$ due to the $\phi$-dependence of $\eta$. In our recent work[9] we have proposed a novel scheme which allows one to obtain an analytical solution to Eq. (1) under the absorbing boundary condition ($c = 0$) at contact ($r = 2a$) and the far-field boundary condition ($c = c_0$) implemented at the hydrodynamic boundary-layer, for an arbitrary direct interaction potential $U(r)$. Now, we can generalize the approach to arbitrary volume fractions, within the effective medium approximation, by solving Eq. (4) with the same scheme used in Ref.[9] The result for the orientation-averaged pair-correlation function is given in the Appendix A. The concentration-dependent binary encounter rate that we obtain reads

$$k_{1,1} = \frac{8\pi[\beta b_{\text{eff}}(\phi)]^{-1}ac_0}{\int_0^{\xi(\phi)} \frac{dx}{\mathcal{G}(x)(x+2)^2} \exp\int_{\xi(\phi)}^x ds(\beta dU/ds + Pe_{\text{eff}}(\phi)\langle\tilde{v}_r^+(\mathbf{r})\rangle)} \tag{5}$$

with $x = (r/a) - 2$, and $\xi(\phi) = \sqrt{(\lambda_D/a)[Pe_{\text{eff}}(\phi)]^{-1}}$, where $\lambda_D$ is the range of the direct interaction (the Debye length in our case). According to the effective medium approach, the effective Peclet number is given by

$$Pe_{\text{eff}}(\phi) = \dot{\gamma}a^2/D_{\text{eff}}(\phi) \tag{6}$$

Furthermore, the orientation-averaged inward (relative) velocity $\langle\tilde{v}_r^+(\mathbf{r})\rangle$ depends uniquely upon the type of flow and its expression is reported in the Appendix A. The integrals in Eq. (5) have to be evaluated numerically. However, as shown in our previous work,[9] the generalized expression for the binary encounter rate can be significantly simplified in the case of an interaction potential which goes through a maximum (potential barrier). This is the case of e.g. the well-known Derjaguin-Landau-Verwey-Overbeek (DLVO) potential,[1] widely used in colloidal science and which is also employed to model the interaction of globular proteins (e.g. lysozime[12]). Applying the steepest-descent method,[9,13] we arrive at the following activated-rate formula for the reaction-limited encounter (aggregation) rate between two Brownian particles in a flowing system at colloid volume fraction $\phi$

$$k_{1,1} \approx 8\pi D_{\text{eff}}(\phi)ac_0[\alpha Pe_{\text{eff}}(\phi) - \beta U_m'']^{1/2} e^{-\beta U_m + 2\alpha Pe_{\text{eff}}(\phi)} \tag{7}$$

$\alpha$ is a coefficient related to the type of flow ($\alpha = 1/3\pi$ in simple shear). $U_m$ is the interaction potential value at the local maximum (i.e. the potential barrier), while $U_m''$ denotes the second derivative evaluated at the maximum.[9] Eq. (7) generalizes Kramers' rate theory[13] to the presence of shear and to concentrated conditions. The associated characteristic time for a binary encounter is given by

$$\tau_{1,1} = \frac{1}{k_{1,1}} \approx \frac{[8\pi D_{eff}(\phi)ac_0]^{-1}}{\sqrt{\alpha Pe_{eff}(\phi) - \beta U_m''}} e^{\beta U_m - 2\alpha Pe_{eff}(\phi)} \qquad (8)$$

It is worth noting that hydrodynamic interactions due to the disturbance of the shear field induced by the relative motion of the particles are accounted for in the analytical treatment as long as the derivation of the rate expression Eq. (5) is concerned [cfr. Eq. (A2) in the Appendix where the role of the hydrodynamic function $A(x)$, originally calculated by Batchelor and Green,[14] is made explicit]. In the subsequent steepest-descent approximation[9] leading to Eqs. (7)-(8) these two-body hydrodynamic interactions are neglected. However, we have checked numerically that the importance of these hydrodynamic corrections is very minor. Indeed the difference between rates calculated using the full expression accounting for them, Eq. (5), and the approximate one, Eq. (7), is practically negligible. This is somewhat expected with DLVO interactions since the interparticle distance range where this hydrodynamic effect is important overlaps with the range where van der Waals attraction dominates. This leads to the hydrodynamic effect being masked (an observation due to Smoluchowski[15]).

Even if the total potential energy of the system can be expressed as a sum over pair-interactions, the force between two particles in a concentrated system is not equal to $-\nabla U(r)$. It is instead given by $-\nabla U_{eff}(r)$, where $U_{eff}$ is the potential of mean force between two particles which contains contributions from the remaining particles.[1,11] For charge-stabilized colloids, these effects have been found to effectively reduce the pair-interaction barrier, $U_m$, with respect to the undisturbed pair-interaction potential between isolated particles.[16-19] In the following we neglect this effect in the comparison with experimental data since in our system $\kappa a = \lambda_D^{-1} a = 24.74$ and $0.19 \leq \phi \leq 0.23$. Hence, the average separation between a particle and its nearest-neighbours is about 1.7-1.8 times the diameter,[20] whereas the range of the screened-Coulomb repulsion is only about 1.04 times the diameter. Further, for a system at equilibrium, the potential of mean force equals $U_{eff}(r) = -k_B T \ln g_{eq}(r)$, where $g_{eq}(r)$ is the pair-correlation function at equilibrium. Under driven, non-equilibrium conditions, the actual pair-correlation function can differ substantially from $g_{eq}(r)$. These additional effects represent a formidable unresolved issue and here are simply omitted.

In order to close the model, we need expressions for the volume fraction-dependent effective viscosity of the medium, $\eta(\phi)$. For hard-spheres, an improved differential viscosity model has been recently proposed by Mendoza and Santamaria-Holek,[21] which yields accurate expressions throughout the entire volume fraction spectrum, from the dilute limit up to close packing, and for both the low and high-shear viscosity. In the case of charge-stabilized particles, the effective viscosity can be further enhanced due to the effective enlargement of the colloid size induced by electrostatics. This effect is especially important at low shear rates where one should care of using expressions for the effective viscosity such as those derived by Russel.[1,22] This effect becomes less and less significant as the shear rate goes up.[22] Since in the following we are going to compare our model predictions with experiments at substantially high shear rates ($\dot{\gamma} \sim 10^3 \, \text{s}^{-1}$) we will neglect this effect and use the viscosity expressions for hard-spheres of Ref.[21].

### III. COMPARISON WITH EXPERIMENTAL DATA

Experimental curves of effective suspension viscosity ($\eta$) as a function of the elapsed shearing time (*t*) are plotted in Fig. 1 for three different volume fractions: $\phi = 0.19$ (a), $\phi = 0.21$ (b), and $\phi = 0.23$ (c). Consistent with previous observations,[4-8] there is a very sharp, explosive upturn in the suspension viscosity due to the onset of self-accelerated aggregation kinetics as soon as the activation-energy, i.e. the argument of the exponential in Eq. (8), vanishes. This happens when on average the formed colloidal clusters reach a shear-activated size, as explained by our theory.[9] A working measure of the characteristic time for aggregation in the experiments ($\tau$) is estimated from the crossing of the asymptotes (according to the protocol introduced by Guery et al.[8]) which is related to the incipient increase of viscosity as a consequence of aggregate formation throughout the system. Rigorously, the theory presented here describes the very initial stage of aggregation where only doublets are formed. However, since the doublets once formed grow further and very quickly to larger aggregates,[9] it is very difficult to monitor the conversion of primary particles to doublets for determining the doublet formation rate. The protocol we used to experimentally estimate the characteristic aggregation time is just the most convenient, systematic procedure to evaluate the time scale of a process which is anyways controlled by the doublet formation rate (since aggregation proceeds very rapidly afterwards due to the larger Peclet values of the aggregates with respect to primary particles). Since the viscosity rise is so fast, the overestimation of the true characteristic aggregation time is certainly not dramatic and represents a systematic effect which does not significantly change the agreement and leaves the scaling with $\dot{\gamma}$ unaltered.

We thus obtain a $\tau_c(\dot{\gamma})$ curve for each $\phi$ investigated. It is seen that, especially at high $\dot{\gamma}$, a modest increase in $\phi$ is able to cause a very significant decrease in the aggregation time. This phenomenon can be attributed to two distinct $\phi$-dependent effects: 1) the effect of collective hydrodynamics which is reflected in a higher effective friction and hence in a higher effective Peclet number; 2) the increase in ionic strength due to increasing the macroion concentration along with $\phi$.

In order to compare predictions from our model with these experimental data, one should consider that the characteristic time of a binary encounter between two particles in the effective medium does not correspond yet to the characteristic aggregation time measured in the experiments. The latter is indeed related to the *total* number of collisions per unit time in the system. Under nondilute conditions, colloidal suspensions exhibit a liquid-like structure where each particle is surrounded by a finite number of nearest-neighbours, corresponding to the first peak in the radial distribution function.[1,23] Hence, in the experiment, all binary collisions of each particle with its nearest-neighbours have the same probability to occur. Thus, the *total* frequency of binary collisions relevant for our comparison is given by the collision frequency between two isolated particles in the effective medium times the total number of equally probable collisions, which is directly proportional to the volume fraction. (This is somewhat analogous to the way the total bombardment frequency of gas molecules on a wall is calculated according to the classical kinetic theory of gases.[24]) The Arrhenius-like form of Eqs. (7)-(8) is helpful in this respect since it allows us to distinguish the collision *frequency*, $\omega_{1,1} = (8\pi D_{eff}(\phi)ac_0)[\alpha Pe_{eff}(\phi) - \beta U_m'']^{1/2}$, i.e. the pre-exponential factor in Eq. (7), from the encounter *efficiency* (the exponential term). Based on these arguments, the total collision frequency in the experimental system (which is relevant to the aggregation time measured), denoted as $\bar{\omega}$, follows upon multiplying the collision frequency for two particles in the effective medium by the total number of (equally probable) collisions in the system. The latter number is given by $zN/2 = 3zV_T\phi/8\pi a^3$, where $z$ is the average number of nearest-neighbours (i.e. of particles in the first coordination shell), and $V_T$ is the total volume of the system. Thus we obtain the following approximate relation for the total binary-collision frequency

$$\bar{\omega} = (3zV_T\phi/8\pi a^3)\omega_{1,1} = (3zV_T\phi D_{eff}(\phi)c_0/a^2)[\alpha Pe_{eff}(\phi) - \beta U_m'']^{1/2} \qquad (9)$$

According to liquid structure theory, the average number of particles in the first shell is always $z \approx 12$ and does not depend much upon the volume fraction.[23] Of course, in the presence of intense shear flow, the local structure is significantly distorted and anisotropic.[11] However, the higher local density in the two upstream quadrants is counterbalanced by particle depletion in the two downstream quadrants. Hence, despite the significant anisotropic shape of the radial distribution function in shear, due to the balancing of densification and depletion in opposite quadrants, the orientation-averaged number of nearest-neighbours in a disordered suspension is not expected to deviate much from the estimate $z = 12$ that we use in our calculations. Thus, the characteristic time for aggregation to be compared with the experimental data is given by

$$\tau \approx \frac{(3zV_T\phi D_{eff}(\phi)c_0/a^2)^{-1}}{\sqrt{\alpha Pe_{eff}(\phi) - \beta U_m''}} e^{\beta U_m - 2\alpha Pe_{eff}(\phi)} \qquad (10)$$

The effect of macroion (colloid) concentration upon the interaction parameters in the model ($U_m$ and $U_m''$), on the other hand, is more difficult to assess. This is due to the impossibility, in the experimental practice, to accurately determine the electrostatic surface potential ($\psi_0$) of the colloidal particles under the nondilute conditions ($\phi \approx 0.2$) of our shearing experiments. In view of this, we have left $\psi_0$ (required in the DLVO calculation of $U_m$ and $U_m''$) as the only adjustable parameter. Once $\psi_0$ is fixed, the DLVO-interaction potential curve for the system can be calculated using the Sader-Carnie-Chan formula valid for high surface electrostatic potentials.[25] The van der Waals attractive component of the DLVO interaction is determined using the standard formula for colloidal spheres as is found in the textbooks[1] with the value of Hamaker constant of the polystyrene-acrylate/water system.

The comparison between theoretical estimates from Eq. (9) and the experimentally measured values of $\tau$, is shown in Fig. 2. For the effective suspension viscosity inside the effective Peclet number we have used the following improved expression for hard-spheres by Mendoza and Santamaría-Holek[21] $\eta(\phi) = \mu[1 - \phi/(1-c\phi)]^{-5/2}$ with $c = (1-\phi_c)/\phi_c$, and $\phi_c = 0.7404$ as prescribed for the high-shear branch. An excellent agreement is found with the following numerical values of the surface potential in the calculation of the DLVO interaction (using the Sader-Carnie-Chan formula[25] for the electric double layer repulsion): $\psi_0 = -45.67$ mV ($\phi = 0.19$), $\psi_0 = -45.64$ ($\phi = 0.21$), and $\psi_0 = -45.60$ ($\phi = 0.23$). A slight decrease of $\psi_0$ with increasing $\phi$ is reasonable since an increase in the macroion concentration (by keeping the salt concentration constant) along with $\phi$ brings about an increase of the total ionic strength of the system. These values of $\psi_0$ are still within the confidence interval of the measured colloid $\zeta$-potential under dilute ($\phi = 5 \times 10^{-4}$) conditions, which was found to be $-45.9 \pm 3$ mV. In measuring the $\zeta$-potential, the same background ionic concentration (17 mM of NaCl) as in the shearing experiments was used. Hence, a value of potential under nondilute conditions which is lower than the value under dilute conditions (with the quantity of added electrolyte being the same) is reasonable.

This comparison demonstrates the capability of the model to capture the volume fraction dependence of reaction-limited aggregation kinetics in shear. Furthermore, the comparison indicates that the major effect behind the significant reduction of the aggregation time upon increasing $\phi$ is the increase in the effective friction and thus in the effective Peclet number which controls the encounter efficiency through the exponential term of Eq. (10). Indeed, $Pe_{eff}$ increases nonlinearly with $\phi$ thus causing a strongly nonlinear increase of the encounter efficiency upon increasing $\phi$. The resulting decrease in the absolute value and increase in the slope of the $\tau_c(\dot{\gamma})$ curve predicted by the theory is qualitatively consistent also with previous experimental data by Guery et al.[8] Finally, the fact that the fitted surface potential is practically constant and changes by less than 0.1mV with $\phi$ suggests that the $\phi$-dependent effect of the ionic strength plays a comparatively minor role.

## V. CONCLUSIONS

We have generalized Kramers' rate theory to the simultaneous presence of shear and concentration effects with application to the shear-induced aggregation rate of charge-stabilized colloids. The results presented here provide insights, for the first time, into the kinetics of reaction-limited (colloidal) aggregation kinetics under shear in concentrated ($\phi > 0.10$) conditions. In order to account for collective hydrodynamics, we have reformulated the Smoluchowski problem with shear by using an effective medium approach. This introduces an effective friction coefficient which depends upon the colloid volume fraction $\phi$ through the effective suspension viscosity. The Smoluchowski equation is then solved using the boundary-condition proposed in recent work.[9] By implementing appropriate expressions for the effective suspension viscosity, we calculated the binary encounter rate for concentrate colloidal suspensions. The theoretical calculations are compared with experimental data of DLVO-interacting colloids (with a potential barrier of approx. 60 $k_BT$) in simple shear at $0.19 \leq \phi < 0.23$. The theory is in good agreement with the experiments and is able to capture the volume fraction dependence of the characteristic aggregation time. Further, the comparison suggests that the main effect behind the significant reduction of the aggregation time upon increasing $\phi$ is to be identified with the nonlinear increase with $\phi$ of the effective Peclet number which in turn causes a strongly nonlinear increase of the collision efficiency. Hence, many-body hydrodynamic interactions play an active role in enhancing the shear-activated barrier-crossing process.[26]

In future work, it is hoped that these findings can be applied to situations directly relevant to biological processes where *in vivo* biochemical reactions between biomolecules occur in crowded and mechanically strained environments.[10]

## APPENDIX A: Derivation of Eq. (5)

Solving Eq. (4) for the pair-correlation function under the far-field boundary-condition proposed in Ref.[9], yields

$$\langle c(\mathbf{r},\phi)\rangle = \left[\exp\int_{\xi(\phi)}^{x} ds(-\beta dU/ds - Pe_{\text{eff}}(\phi)\langle \tilde{v}_r^+(\mathbf{r})\rangle)\right]$$
$$\times \left\{ c_0 + \frac{G}{8\pi a[\beta b_{\text{eff}}(\phi)]^{-1}} \int_{\xi(\phi)}^{x} \frac{ds}{\mathcal{G}(s)(s+2)^2} \right.$$
$$\left. \times \exp\int_{\xi(\phi)}^{x} ds(\beta dU/ds + Pe_{\text{eff}}(\phi)\langle \tilde{v}_r^+(\mathbf{r})\rangle) \right\} \quad (A.1)$$

Under *dilute* conditions ($\phi \to 0$) the nondimensionalized hydrodynamic boundary-layer thickness is given by $\xi = \delta/a = \sqrt{(\delta/a)/Pe}$.[9] Under *nondilute* conditions, $\xi$ becomes as well a function of the volume fraction, $\xi(\phi) = \sqrt{(\lambda/a)[Pe_{\text{eff}}(\phi)]^{-1}}$. With simple shear, the flow term $\langle \tilde{v}_r^+(\mathbf{r})\rangle$ is given by

$$\langle \tilde{v}_r^+(\mathbf{r})\rangle \equiv -(1/3\pi)(x+2)[1-A(x)] \quad (A.2)$$

$A(x)$ is the hydrodynamic retardation function for two particles approaching each other and its expression can be found in the textbooks.[1] The solution to the Smoluchowski problem with shear, Eq. (4), is completely determined upon implementing the absorbing boundary-condition at contact which identifies the inward flux $G$ of particles toward the tagged one as

$$G = \frac{8\pi[\beta b_{\text{eff}}(\phi)]^{-1} a c_0}{2 \int_0^{\xi(\phi)} \frac{dx}{\mathcal{G}(x)(x+2)^2} \exp\int_{\xi(\phi)}^{x} ds(\beta dU/ds + Pe_{\text{eff}}(\phi)\langle \tilde{v}_r^+(\mathbf{r})\rangle)} \quad (A.3)$$

Eq. (5) readily follows upon applying $k_{1,1} \equiv 2G$.[27]

## APPENDIX B: Experimental Section

The colloidal system used to perform the experiments reported here is a surfactant-free colloidal dispersion in water, constituted by styrene-acrylate copolymer particles supplied by BASF SE (Ludwigshafen, Germany). The nearly monodisperse particles have mean radius of $a = 60 \pm 1$ nm, and were characterized by both dynamic and static light scattering (using a BI-200 SM goniometer system, Brookhaven Instruments, NY). In order to avoid contamination, a thorough cleaning of the suspensions by mixing with an ion-exchange resin (Dowex MR-3, Sigma-Aldrich) was performed. To check that the suspensions were free of impurities after the cleaning procedure, we measured the surface tension by means of the Wilhelmy plate method with a DCAT-21 tensiometer (Dataphysics, Germany) and only suspensions with surface tension $\geq 71.7$ mN/m were used for the investigations. For the shearing experiments, a small amount of electrolyte (NaCl) was added to make up the ionic background. In fact, with de-ionized suspensions, the shearing time at which viscosity rises due to aggregation would be very long (on the order of days). This can seriously affect the system and the reproducibility of the experiments due to solvent evaporation. However, the final NaCl concentration in the sample (17mM) is well below the critical coagulation concentration (50mM with NaCl). The long-time stability of each suspension after adding the NaCl solution was checked by light scattering. To induce the shear flow under shear-rate control and to simultaneously measure the viscosity of the flowing suspension, a strain-controlled ARES rheometer (Advanced Rheometric Expansion System, TA Instruments, Germany) with Couette geometry has been employed. The gap between the outer cylinder and the inner one is 1 mm and the diameter of the latter is 34 mm. The outer cylinder is temperature controlled at $T = 298 \pm 0.1\,\text{K}$ and, in order to prevent evaporation, a solvent trap has been fixed on the outer rotating cylinder. In all the experiments we used deionized water (milli-Q, Millipore) and the mixing of the latex suspensions with NaCl solutions was done in such a way to avoid heterogeneities in the concentration field which could cause the aggregation kinetics to speed up in locally more concentrated regions. It is worth noting that the sampling of all the NaCl-solution/latex mixtures was done carefully with a top-cut pipette in order to avoid any local shearing during the sampling that could induce aggregation. In order to ensure reproducibility, each time the shearing was switched on 7 minutes from the time of mixing between latex and background NaCl solution. For each point in the $\phi$ - $\dot{\gamma}$ plane investigated, at least three repetitions were done. The experimental error on the aggregation time $\tau$ has never been found to exceed 15% of the mean value. $\zeta$-potential measurements were carried out using a Zetasizer Nano instrument (Malvern, UK), on dilute suspensions ($\phi = 5 \times 10^{-4}$) at the same ionic background concentration (17mM NaCl) used for the shearing experiments.

## ACKNOWLEDGMENTS

BASF SE (Ludwigshafen, Germany) is gratefully acknowledged for supplying the colloidal particles. Financial support from the Swiss National Science Foundation (grant No. 200020-126487/1) is gratefully acknowledged.

**FIGURE 1**

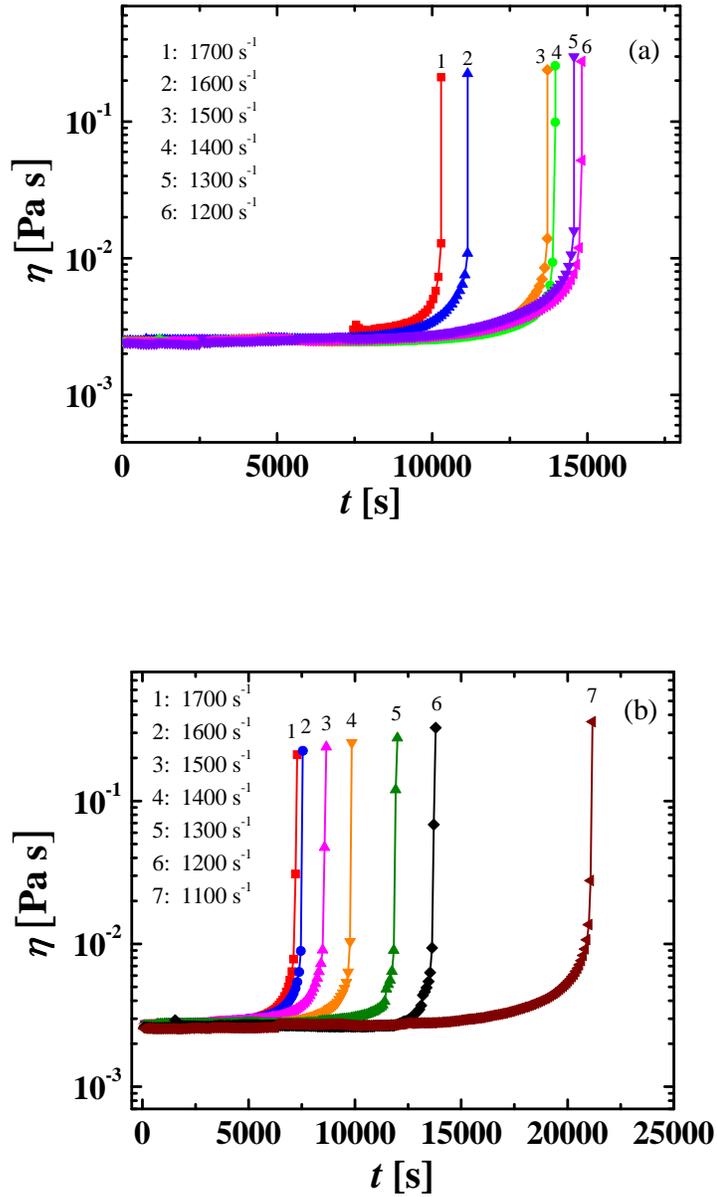

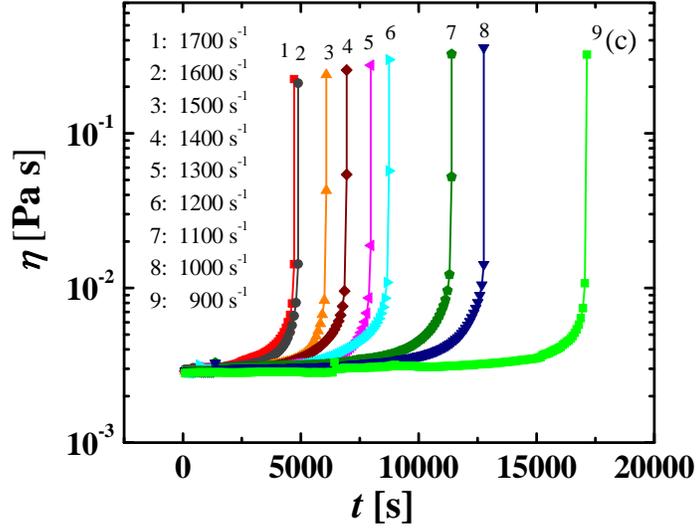

**FIGURE 2**

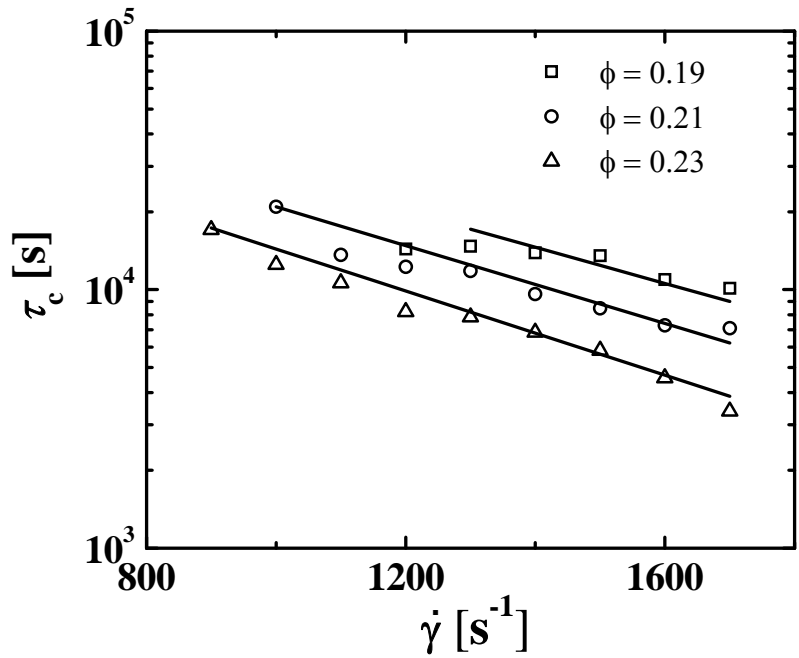

**FIGURE CAPTIONS**

FIG. 1. Suspension viscosity as a function of the shearing time under steady shear for charge-stabilized colloids at volume fractions $\phi = 0.19$ (a), $\phi = 0.21$ (b), and $\phi = 0.23$ (c), and at a varying shear rate $\dot{\gamma}$ (see legends). Added electrolyte for all conditions: 17 mM (NaCl). The characteristic time of aggregation ($\tau$) is estimated as shown schematically in (a).

FIG. 2. Characteristic aggregation time under nondilute conditions (see legend) as a function of the applied shear-rate. Symbols: data points from the experiments reported in Figure 1. Solid lines: theoretical calculations using Eq. (10) (see Text).